\documentclass[twocolumn]{revtex4}
\bibpunct{[}{]}{;}{a}{,}{,}%

\usepackage{graphicx}

\begin{document}

\title{Zakharov simulation study of spectral features of on-demand Langmuir turbulence in an inhomogeneous plasma}

\author{B. Eliasson}
\affiliation{Department of Physics, Ume{\aa} University, SE-901 87
Ume{\aa}, Sweden} \affiliation{Theoretische Physik IV,
Ruhr--Universit\"at Bochum, D-44780 Bochum, Germany}

\author{B. Thid\'e}
\affiliation{ Swedish Institute of Space Physics, P. O. Box 537,
SE-751 21 Uppsala, Sweden } \affiliation{LOIS Space Centre,
V\"axj\"o University, SE-351 95 V\"axj\"o, Sweden}

\begin{abstract}
We have performed a simulation study of Langmuir turbulence in the
Earth's ionosphere by means of a Zakharov model with parameters
relevant for the F layer. The model includes dissipative terms to
model collisions and Landau damping of the electrons and ions, and a
linear density profile, which models the ionospheric plasma
inhomogeneity whose length scale is of the order $10$--$100$ km. The
injection of energy into the system is modeled by a constant source
term in the Zakharov equation. Langmuir turbulence is excited
``on-demand'' in controlled ionospheric modification experiments
where the energy is provided by an HF radio beam injected into the
overhead ionospheric plasma.  The ensuing turbulence can be studied
with radars and in the form of secondary radiation recorded by
ground-based receivers. We have analyzed spectral signatures of the
turbulence for different sets of parameters and different altitudes
relative to the turning point of the linear Langmuir mode where the
Langmuir frequency equals the local plasma frequency. By a
parametric analysis, we have derived a simple scaling law, which
links the spectral width of the turbulent frequency spectrum to the
physical parameters in the ionosphere. The scaling law provides a
quantitative relation between the physical parameters (temperatures,
electron number density, ionospheric length scale, etc.) and the
observed frequency spectrum. This law may be useful for interpreting
experimental results.
\end{abstract}

\maketitle
%
%

%


%
%

\section{Introduction}
The interaction between an electromagnetic wave and an inhomogeneous
plasma layer leads to phenomena on different temporal and spatial
scales. Laboratory experiments \citep{Kim74} have demonstrated
cavity formation and trapping of radio frequency (RF) electrostatic
fields due to density modification of the plasma near the critical
density where the plasma frequency equals that of the electrostatic
RF pump field. Ionospheric cavitons were observed near the critical
layer of a high-power radio wave injected into the overhead
ionosphere \citep{Wong87}. In these experiments, both
decimeter-scale cavitons due to the ponderomotive force by
mode-converted electrostatic waves (electrostatic cavitons) and
kilometer-sized cavitons due to the thermal pressure of the heated
electrons (thermal cavitons) were observed.

By analyzing the frequency spectrum of the electromagnetic waves
recorded by ground-based receivers, it was found experimentally in
Troms{\o} that strong, systematic, structured, wide-band secondary
HF radiation escapes from the turbulent interaction region
\citep{Thide82}. This and other observations demonstrated that
complex interactions, including weak and strong EM turbulence, are
involved \citep{Leyser01,Carozzi02}. Naturally enhanced ion acoustic
lines (NEIALs) observed by several radar facilities
\citep{Foster88,Sedgemore01} have in some cases been interpreted as
the result of Langmuir turbulence
\citep{Sedgemore01,Forme93,Guio06}. Several simulation studies in
homogeneous plasmas have demonstrated the basic processes of
Langmuir turbulence and wave collapse
\citep{Nicholson84,DuBois91,DuBois93,Hansen92,Robinson97,Mjolhus03}.
Simulation studies with an inhomogeneous Zakharov equation, where
the electromagnetic field was modeled by an external dipolar field,
demonstrated density modification and the excitation of localized
electric fields and ion-acoustic oscillations in the plasma
\citep{Morales74,Morales77}. Resonant absorption of electromagnetic
waves and electrostatic turbulence was studied experimentally and
numerically \citep{Cros01} with a modified Zakharov model and
demonstrated the density modification and transition to turbulence
for different amplitudes of the electromagnetic RF pump wave.

In the present article, we present results from long-time series
simulations of the driven Zakharov system of equations in order to
investigate the statistical properties of Langmuir turbulence in an
inhomogeneous plasma, and to obtain the dependence of the spectrum
on physical parameters. The manuscript is organized in the following
fashion. In Section 2, we present the mathematical model in the form
of the driven Zakharov equations for an inhomogeneous plasma
background, including Landau damping and collision terms. The
numerical setup and results are presented in Section 3, where
frequency spectra and their dependence on physical variables are
discussed, and an approximate scaling law for the spectral width is
derived. The results obtained are summarized in Section 4.

\section{The inhomogeneous Zakharov equation}

In the Zakharov model, the Langmuir wave field in the presence of a
linear density gradient, a constant dipole pump field and slowly
varying density fluctuations is governed by
\citep{Morales74,Morales77}
\begin{equation}
\frac{2i}{\omega_{pe}}\frac{\partial E}{\partial
t}-\left(\frac{x}{L}
+\frac{n_s}{n_0}-i\frac{\gamma_L}{\omega_{pe}}\right)E+3\lambda_{De}^2
\frac{\partial^2E}{\partial x^2}=E_{\rm pump}, \label{eq1}
\end{equation}
where $i$ is the imaginary unit, $\omega_{pe}=(n_0e^2/\varepsilon_0
m_e)^{1/2}$ is the electron plasma frequency, $x$ is the vertical
spatial coordinate, $L$ is the local length scale of the ionospheric
profile, $n_s$ is the slowly varying electron density perturbation
due to ion-acoustic fluctuations, $n_0$ is the equilibrium electron
number density, $\gamma_L$ is the collision frequency,
$\lambda_{De}=v_{Te}/\omega_{pe}$ is the electron Debye length,
$v_{Te}=(k_B T_e/m_e)^{1/2}$ is the electron thermal speed, $e$ is
the magnitude of the electron charge, $m_e$ is the electron mass,
$\varepsilon_0$ is the vacuum permittivity, $k_B$ is Boltzmann's
constant, and $T_e$ is the electron temperature. The source term
$E_{\rm pump}$ represents the energy provided by electromagnetic
waves \citep{Thide82}. If the radio wave has an oblique incidence to
the plasma gradient, then it will be reflected at an altitude
slightly lower than the critical height where the pump frequency
equals the plasma frequency. A small fraction of the electromagnetic
wave will tunnel up to the critical height where it excites
large-amplitude electrostatic waves. In our model, the dipole pump
represents the evanescent electromagnetic wave that has tunneled up
to the critical altitude. Other mode conversion processes may be
important due to the presence of the geomagnetic field
\citep{Mjolhus90}, which we have neglected here. The ion acoustic
fluctuations in the presence of the ponderomotive force are governed
by
\begin{equation}
  \frac{\partial^2 n_s}{\partial t^2}+2\gamma_s\frac{\partial n_s}{\partial t}
  -C_s^2\frac{\partial^2 n_s}{\partial x^2}
  =\frac{\varepsilon_0}{4 m_i}\frac{\partial^2|E|^2}{\partial x^2}
  \label{eq2}
\end{equation}
where $C_s=[k_B(T_e+3T_i)/m_i]^{1/2}$ is the ion acoustic speed,
$T_i$ is the ion temperature, $m_i$ is the ion mass, and $\gamma_s$
is the ion collision frequency \citep{Nicholson84} due to collisions
and/or Landau damping, which, for simplicity, we have taken to be
constant.
\section{Numerical results}

\begin{figure}[htb]
  \centering
  \includegraphics[width=8.5cm]{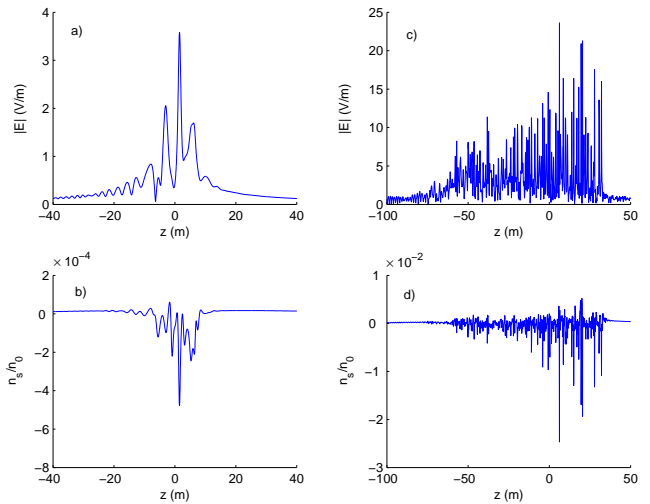}
  \caption{a) Profiles of the amplitude of Langmuir electric field, $|E|$ and b) the
  electron number density perturbation $n_s/n_0$ for $E_{\rm pump}=1.0\times 10^{-4}\,\mathrm{V/m}$.
  c) and d): The same as a) and b) but for $E_{\rm pump}=1.0\times 10^{-3}\,\mathrm{V/m}$.
  The other parameters are $L=50\,\mathrm{km}$,
  $\gamma_{L}=1.0\times 10^3\,\mathrm{s}^{-1}$, and
  $\gamma_{s}=1.0\times 10^3\,\mathrm{s}^{-1}$.
  }
  \label{Fig1}
\end{figure}

\begin{figure}[htb]
  \centering
  \includegraphics[width=8.5cm]{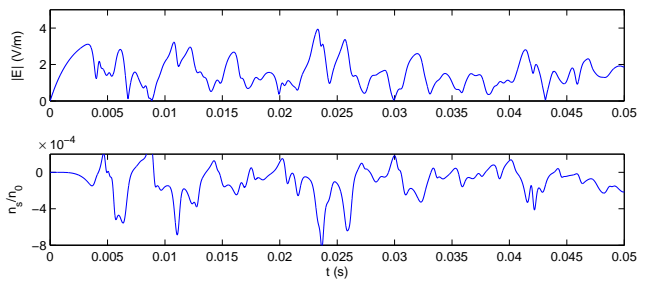}
  \caption{The time-dependent amplitude at $z=0$ of the Langmuir electric field $|E|$ (upper panel)
  and of the electron number density perturbation $n_s/n_0$ (lower panel) for $E_{\rm
  pump}=1.0\times 10^{-4}\,\mathrm{V/m}$.
  The other parameters are $L=50\,\mathrm{km}$,
  $\gamma_{L}=1.0\times 10^3\,\mathrm{s}^{-1}$, and
  $\gamma_{s}=1.0\times 10^3\,\mathrm{s}^{-1}$.
  }
  \label{Fig2}
\end{figure}

\begin{figure}[htb]
  \centering
  \includegraphics[width=8.5cm]{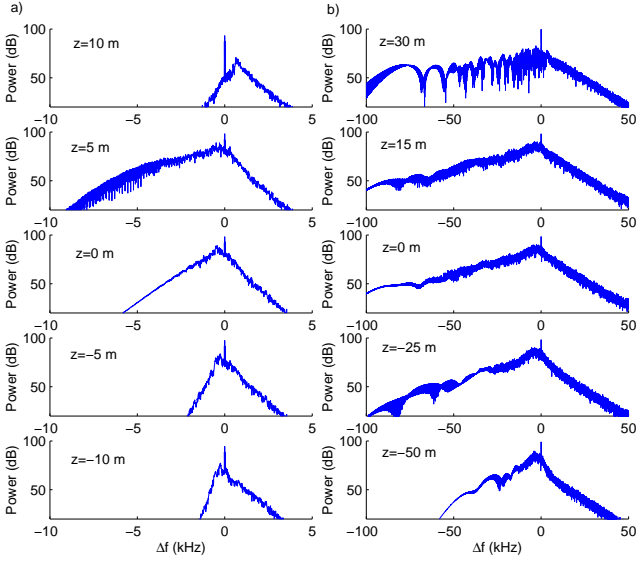}
  \caption{Frequency spectra (in dB) of $E$ at different altitudes, a)
  $z=10\,\mathrm{m}$,
  $z=5\,\mathrm{m}$,
  $z=0\,\mathrm{m}$,
  $z=-5\,\mathrm{m}$, and
  $z=-10\,\mathrm{m}$,
   (top to bottom
  panel), for $E_{\rm pump}=10^{-4}\,\mathrm{V/m}$ and b)
  $z=30\,\mathrm{m}$,
  $z=15\,\mathrm{m}$,
  $z=0\,\mathrm{m}$,
  $z=-25\,\mathrm{m}$, and
  $z=-50\,\mathrm{m}$, (top to bottom
  panel), for $E_{\rm pump}=10^{-3}\,\mathrm{V/m}$.
  The other parameters are $L=50\,\mathrm{km}$,
  $\gamma_{L}=1.0\times 10^3\,\mathrm{s}^{-1}$, and
  $\gamma_{s}=1.0\times 10^3\,\mathrm{s}^{-1}$.
  }
  \label{Fig3}
\end{figure}

In order to investigate the nonlinear dynamics and spectral features
of the Zakharov system, we have solved Eqs.~(\ref{eq1}) and
(\ref{eq2}) for different sets of parameters. We have used
$n_0=5\times 10^{11}\,\mathrm{m}^{-3}$, $T_e=T_i=2000\,\mathrm{K}$,
and we have used $m_i=26.8\times10^{-27}\,\mathrm{kg}$ (oxygen ions)
in our simulations, so that $\omega_{pe}=4.0\times
10^7\,\mathrm{s}^{-1}$, $v_{Te}=1.7\times10^5\,\mathrm{m/s}$ and
$C_s=2.0\times10^3\,\mathrm{m/s}$. The pump frequency, which is set
equal to the local plasma frequency, is
$f_0=\omega_{pe}/2\pi=6.34\,\mathrm{MHz}$. In the numerical
procedure, the Zakharov system was advanced in time with a standard
fourth-order Runge-Kutta scheme with a timestep of
$10^{-6}\,\mathrm{s}$. The spatial derivatives were approximated
with centered, second-order difference approximations with a spatial
grid spacing of $5\,\mathrm{cm}$. For weaker pump fields $E_{\rm
pump}$ of order $10^{-4}\,\mathrm{V/m}$, we used the spatial domain
$-100\,\mathrm{m}\leq z \leq 100\,\mathrm{m}$, and for the stronger
pump fields of order $10^{-3}\,\mathrm{V/m}$, we used the spatial
domain $-200\,\mathrm{m}\leq z \leq 200\,\mathrm{m}$.

 In Fig.~\ref{Fig1}, we show the profiles of the electric field and the
density fluctuations at the end of the simulation, for two different
values of the pump field. For a pump field of $E_{\rm
pump}=0.1\mathrm{mV/s}$ [panels a) and b)], we see an excited
electrostatic field of the order $|E|\sim 3\mathrm{V/m}$ correlated
with a density depletion of $n_s/n_0\sim -4\times10^{-4}$. The
turbulent field is localized in a relatively small interval of $z$
between $-10\,\mathrm{m}$ and $10\,\mathrm{m}$, correlated with
electric field maxima are density minima. For the larger pump field
$E_{\rm pump}=1.0\,\mathrm{V/m}$, displayed in panels c) and d), the
turbulence takes place in a larger altitude interval, between
$z\simeq -60\,\mathrm{m}$ and $z\simeq 30\,\mathrm{m}$. Here, the
Langmuir field is larger, of the order $10$--$20\,\mathrm{V/m}$, and
the maximum density depletion, correlated with electric field
maxima, is of the order $n_s/n_0\sim -10^{-2}$.

\begin{figure}[htb]
  \centering
  \includegraphics[width=8.5cm]{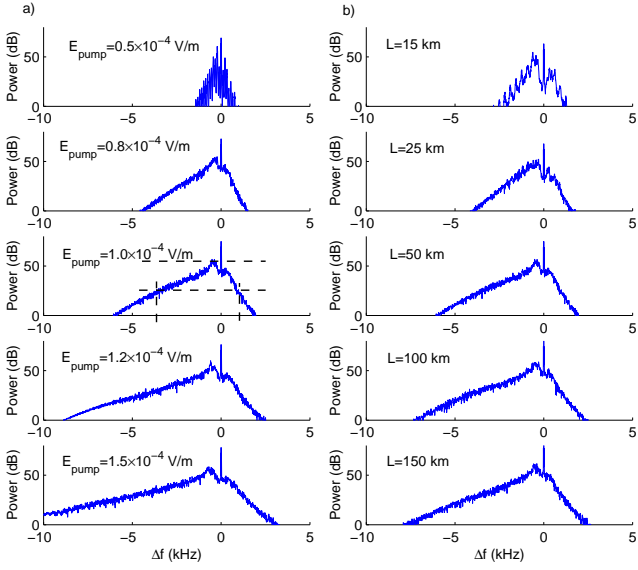}
  \caption{a) Frequency spectra (in dB) of $E$ (spatially averaged) for different pump
  amplitudes $E_{\rm pump}=0.5\times 10^{-4}\,\mathrm{V/m}$,
  $E_{\rm pump}=0.8\times 10^{-4}\,\mathrm{V/m}$, $E_{\rm pump}=1.0\times 10^{-4}\,\mathrm{V/m}$,
  $E_{\rm pump}=1.2\times 10^{-4}\,\mathrm{V/m}$, and $E_{\rm pump}=1.5\times 10^{-4}\,\mathrm{V/m}$ (top to bottom
  panel), for $L=50\,\mathrm{km}$. The spectral width is defined as the frequency difference between the
high-frequency and low-frequency points where the power has dropped
30 dB compared to the downshifted maximum, as indicated with dashed
lines for the case $E_{\rm pump}=1.0\times 10^{-4}$ V/m; here the
spectral width is approximately 5 kHz.
  b) Frequency spectra (in dB) of $E$ (spatially averaged) for different
  ionospheric length scales $L=15\,\mathrm{km}$, $L=25\,\mathrm{km}$, $L=50\,\mathrm{km}$, $L=100\,\mathrm{km}$,
  and $L=150\,\mathrm{km}$ (top to bottom
  panel), for $E_{\rm pump}=1.0\times 10^{-4}\,\mathrm{V/m}$. The other
  parameters are $\gamma_{L}=1.0\times 10^3\,\mathrm{s}^{-1}$ and $\gamma_{s}=1.0\times 10^3\,\mathrm{s}^{-1}$
  }
  \label{Fig4}
\end{figure}

\begin{figure}[htb]
  \centering
  \includegraphics[width=8.5cm]{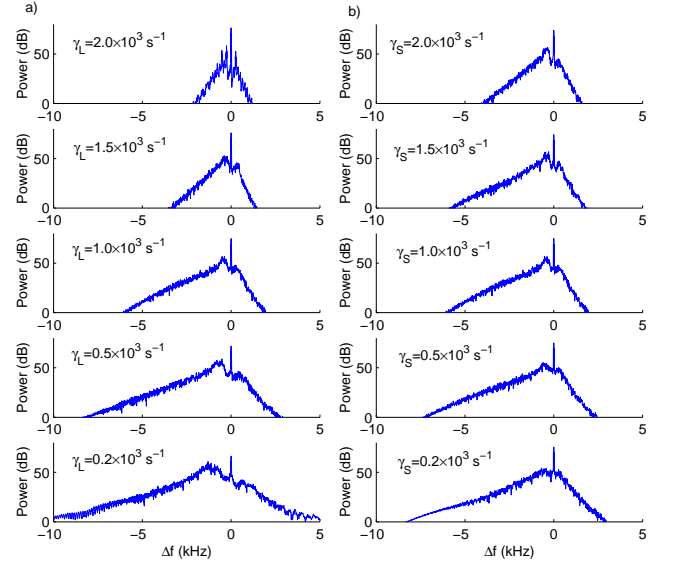}
  \caption{a) Frequency spectra (in dB) of $E$ (spatially averaged) for electron collision frequencies
  $\gamma_{L}=2.0\times 10^3\,\mathrm{s}^{-1}$,
  $\gamma_{L}=1.5\times 10^3\,\mathrm{s}^{-1}$,
  $\gamma_{L}=1.0\times 10^3\,\mathrm{s}^{-1}$,
  $\gamma_{L}=0.5\times 10^3\,\mathrm{s}^{-1}$, and
  $\gamma_{L}=0.2\times 10^3\,\mathrm{s}^{-1}$ (top to bottom
  panel), for $\gamma_{s}=1.0\times 10^3\,\mathrm{s}^{-1}$.
  b) Frequency spectra (in dB) of $E$ (spatially averaged) for different
  ion collision frequencies   $\gamma_{L}=2.0\times 10^3\,\mathrm{s}^{-1}$,
  $\gamma_{s}=1.5\times 10^3\,\mathrm{s}^{-1}$,
  $\gamma_{s}=1.0\times 10^3\,\mathrm{s}^{-1}$,
  $\gamma_{s}=0.5\times 10^3\,\mathrm{s}^{-1}$, and
  $\gamma_{s}=0.2\times 10^3\,\mathrm{s}^{-1}$ (top to bottom
  panel), for $\gamma_{L}=1.0\times 10^3\,\mathrm{s}^{-1}$.
  The other parameters are $L=50\,\mathrm{km}$
  and $E_{\rm pump}=1.0\times 10^{-4}\,\mathrm{V/m}$.
  }
  \label{Fig5}
\end{figure}

\begin{figure}[htb]
  \centering
  \includegraphics[width=8.5cm]{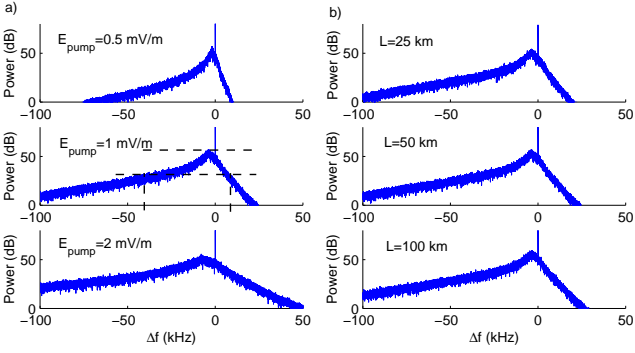}
  \caption{a) Frequency spectra (in dB) of $E$ (spatially averaged) for different pump amplitudes
  $E_{\rm pump}=0.5\times 10^{-3}\,\mathrm{V/m}$,
  $E_{\rm pump}=1.0\times 10^{-3}\,\mathrm{V/m}$,
  and $E_{\rm pump}=2\times 10^{-3}\,\mathrm{V/m}$ (top to bottom
  panel), for $L=50\,\mathrm{km}$. We have indicated the spectral width
  with dashed lines, in the same manner as in Fig.~\ref{Fig4}, for the case
  $E_{\rm pump}=1.0\times10^{-3}$ V/m; here the spectral width is approximately $50$ kHz.
  b) Frequency spectra (in dB) of $E$ (spatially averaged) for different
  ionospheric length scales
  $L=25\,\mathrm{km}$,
  $L=50\,\mathrm{km}$,
  and $L=100\,\mathrm{km}$ (top to bottom
  panel), for $E_{\rm pump}=1.0\times 10^{-3}\,\mathrm{V/m}$.
  The other parameters are $\gamma_{L}=1.0\times 10^3\,\mathrm{s}^{-1}$ and $\gamma_{s}=1.0\times 10^3\,\mathrm{s}^{-1}$.
  }
  \label{Fig6}
\end{figure}

\begin{figure}[htb]
  \centering
  \includegraphics[width=8.5cm]{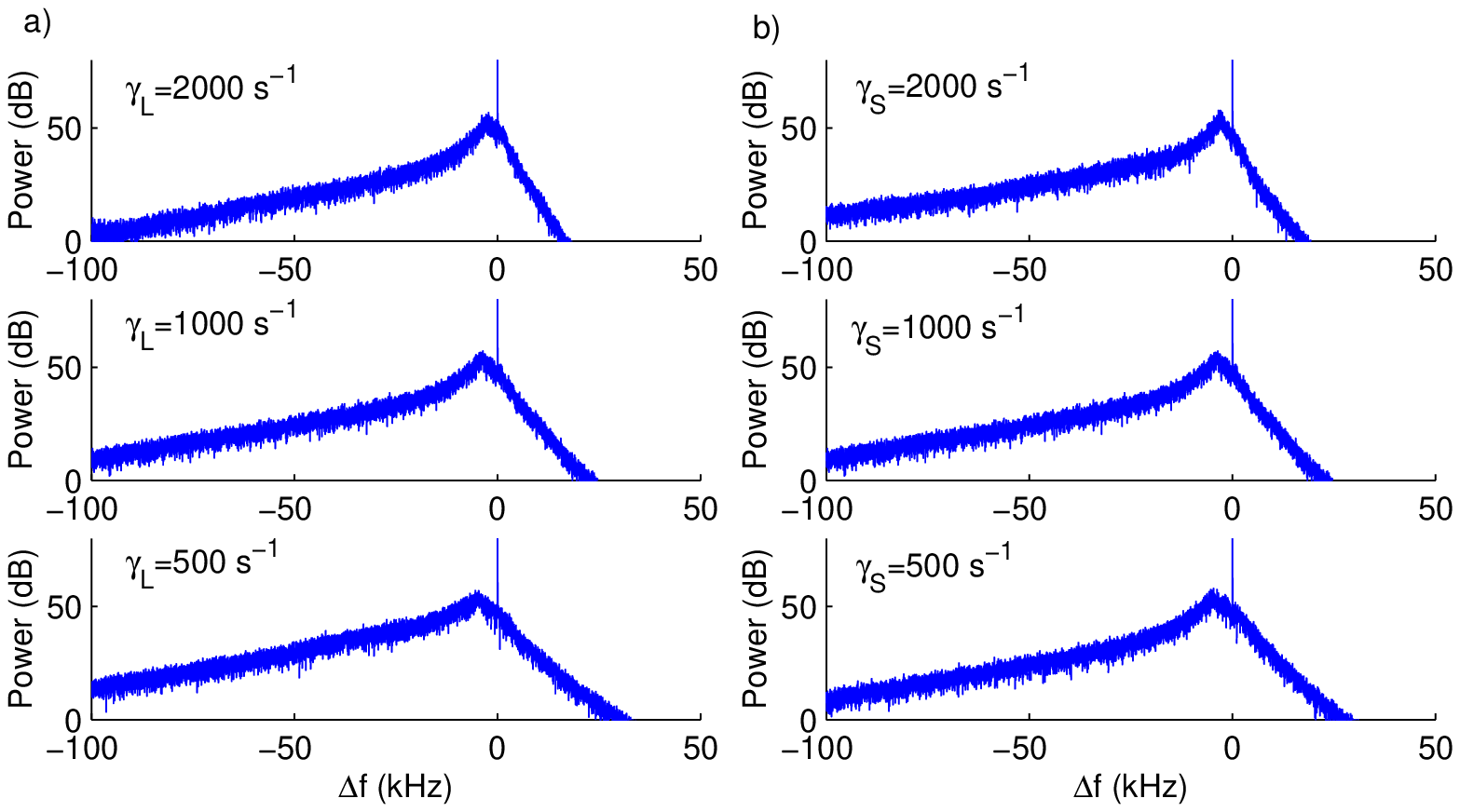}
  \caption{a) Frequency spectra (in dB) of $E$ (spatially averaged) for electron collision frequencies
  $\gamma_{L}=2.0\times 10^3\,\mathrm{s}^{-1}$,
  $\gamma_{L}=1.0\times 10^3\,\mathrm{s}^{-1}$, and
  $\gamma_{L}=0.5\times 10^3\,\mathrm{s}^{-1}$, (top to bottom
  panel), for $\gamma_{s}=1.0\times 10^3\,\mathrm{s}^{-1}$.
  b) Frequency spectra (in dB) of $E$ (spatially averaged) for different
  ion collision frequencies
  $\gamma_{s}=2.0\times 10^3\,\mathrm{s}^{-1}$,
  $\gamma_{s}=1.0\times 10^3\,\mathrm{s}^{-1}$, and
  $\gamma_{s}=0.5\times 10^3\,\mathrm{s}^{-1}$,
  (top to bottom panel), for $\gamma_{L}=1.0\times
  10^3\,\mathrm{s}^{-1}$.
  The other parameters are $L=50\,\mathrm{km}$
  and $E_{\rm pump}=1.0\times 10^{-3}\,\mathrm{V/m}$.
  }
  \label{Fig7}
\end{figure}

The time dependent amplitude of the Langmuir field and the density
fluctuations at $z=0$ are shown in Fig.~\ref{Fig2}, for the first 50
ms (physical time). After an initial growth phase of the fields
during the first 3 ms, the fluctuations enter a steady-state
turbulent phase during the rest of the time. In Fig.~\ref{Fig3}, we
are analyzing the frequency spectrum of the electric field at
different altitudes. In order to obtain the frequency spectrum at
each altitude, the simulation was run for $500\,\mathrm{ms}$. The
time series of the electric field was subdivided into five
time-slices of $100\,\mathrm{ms}$, and each time slice is Fourier
transformed in time with a Hamming window. Finally, the average is
taken of the five power spectra. The averaged power spectrum (in dB)
is visualized in Fig.~\ref{Fig3} for $E_{\rm
pump}=0.1\,\mathrm{mV/m}$ and $E_{\rm pump}=1\,\mathrm{mV/m}$, at
different altitudes. We see that especially the the down-shifted
part of the power spectrum differs significantly between different
altitudes, while the upshifted part of the spectra vary in their
amplitudes but their logarithmic slopes are almost the same at
different altitudes. In column a) of Fig.~3, we see that the
spectrum is much wider at $z=5$~m than at $z=-5$~m. Hence, {\it in
situ} measurements by satellites or rockets would be very sensitive
to the exact location of the measuring probe. In a thought
ground-based experiment, the electrostatic turbulence converts to
electromagnetic radiation by some process \citep{Stubbe84}, and the
escaping radiation is monitored by receiving equipment of the
ground. The conversion process may involve the mode-conversion of
the electrostatic waves against density gradients, resulting in a
mixture of frequency components from different altitudes in the
resultant electromagnetic radiation. The radiation field from an
extended body can be expressed in terms of an integral over the
retarded current density. For electrostatic turbulence, the current
density and electrostatic field are linked by the Maxwell equation
$\partial {\bf E}/\partial t=-{\bf j}/\varepsilon_0$, and hence the
radiation field is proportional to the spatial integral of the
electrostatic field over the turbulent region. We here do not model
in detail the conversion process but will assume that the escaping
radiation contains frequency components from the spatial average of
the electrostatic field (possibly modified by the propagation
characteristics of the ionospheric plasma). This model may be valid
as long as the wavelengths of the electrostatic waves are much
smaller than that of the vacuum electromagnetic waves. In our case
the electrostatic waves are on decimeter or meter scale, while the
electromagnetic waves are $\sim 60\,\mathrm{m}$. Hence, we have
taken the spatial average of the complex electric field $E$ over the
simulation box. The spatially averaged electric field is Fourier
analyzed in a similar manner as in Fig.~\ref{Fig3}, and the
resulting frequency spectra are plotted in
Figs.~\ref{Fig4}--\ref{Fig7} for different sets of parameters.

In principle, the eight physical parameters that can be varied are
$E_{\rm pump}$, $L$, $\gamma_{L}$, $\gamma_{s}$, $T_{ e}$, $T_{i}$,
$n_0$, and $m_i$. However, the number of parameters can be reduced
by normalizing the system, in the spirit of, e.g.,
\citet{Morales77}. Introducing dimensionless, primed variables
according to
\begin{equation}
  \label{scaling_t}
  t=({3v_{Te}^2}/{2C_s^2\omega_{pe}}) t',
\end{equation}
\begin{equation}
  x=({3v_{Te}^2}/{2C_s\omega_{pe}}) x',
\end{equation}
\begin{equation}
  n_s=({4C_s^2n_0}/{3v_{Te}^2}) n_s',
\end{equation}
\begin{equation}
  E=({4C_s}/{v_{Te}}) ({n_0m_iC_s^2}/{3\varepsilon_0})^{1/2} E',
\end{equation}
\begin{equation}
  \label{scaling_Epump}
  E_{\rm pump}=({16C_s^3}/{3v_{Te}^3}) ({n_0m_iC_s^2}/{3\varepsilon_0})^{1/2} E_{\rm
  pump}',
\end{equation}
\begin{equation}
  L=({9v_{Te}^4}/{8C_s^3\omega_{pe}}) L',
\end{equation}
\begin{equation}
  \gamma_L=({4C_s^2\omega_{pe}}/{3v_{Te}^2}) \gamma_L',
\end{equation}
\begin{equation}
  \gamma_s=({C_s^2\omega_{pe}}/{3v_{Te}^2}) \gamma_s',\\
  \label{scaling_gs}
\end{equation}
the Zakharov system (\ref{eq1})--(\ref{eq2}) is transformed into
\begin{equation} \label{tot_scaled1}
   i \frac{\partial E'}{\partial t'}-\left(\frac{x'}{L'}+n_s'-i\gamma_L'\right)
  E'+\frac{\partial^2 E'}{\partial x'^2}=E_{\rm pump}'
\end{equation}
\begin{equation}
  \frac{\partial^2 n_s'}{\partial t'^2}+\gamma_s'\frac{\partial n_s'}{\partial t'}
  -\frac{\partial^2 n_s'}{\partial x'^2}
  =\frac{\partial^2 |E'|^2}{\partial x'^2},
  \label{primed}
\end{equation}
and we see that the scaled system have only four free parameters
$L'$, $\gamma_L'$, $E_{\rm pump}'$, and $\gamma_s'$. We can state
the similarity principle that for any change of parameters
(temperatures, densities, etc.) in the dimensional system such that
the dimensionless parameters $L'$, $\gamma_L'$, $E_{\rm pump}'$, and
$\gamma_s'$ are unchanged, the system is {\emph self-similar}, i.e.,
it will have the same solutions as the original system up to linear
scalings. We have chosen to vary only the four parameters $E_{\rm
pump}$, $L$, $\gamma_{L}$ and $\gamma_{s}$ in our numerical
simulations, while keeping the other parameters constant.

\begin{figure}[htb]
  \centering
  \includegraphics[width=8.5cm]{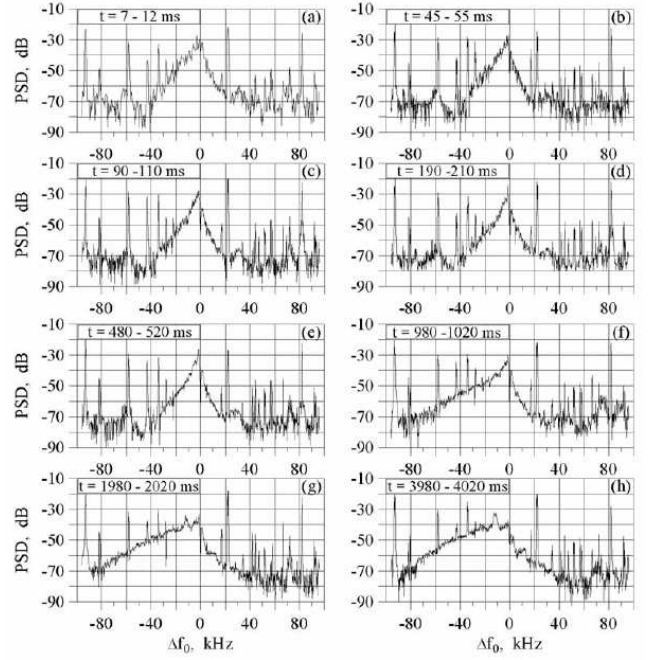}
  \caption{
  a) SEE spectra recorded at the Sura
  heating facility in Nizhny Novgorod, Russia, on 25 September 1991
  at 13:30--13:50 LT. The spectra are
  measured at different times (indicated in the boxes) after the
  switch-on of the pump wave, in a 10 s on, 50 s off mode with pump
  frequency 5828 MHz and ERP power 30 MW. The spectra are primarily downshifted,
  and a slow modification of the
  spectra can be seen, with wider spectra at later times.
  The narrow spikes are due to interfering radio stations.
  (After \citet{Frolov04}.)
  }
  \label{Fig8}
\end{figure}

All the spectra in Figs.~\ref{Fig4}--\ref{Fig7} have some common
features. The spectra are asymmetric, where the downshifted part of
the frequency spectrum is about three times wider than the upshifted
one, although the relative width seems to increase for the
downshifted spectrum for larger pump amplitude; see Figs.~\ref{Fig6}
and \ref{Fig7}. We also see a downshifted maximum in the spectra,
whose downshift varies from a few kHz up to $\sim 5\,\mathrm{kHz}$,
depending on the parameters. These general spectral features are
frequently observed in experiments
\citep{Thide82,Thide89,Thide90,Frolov04}. A few examples of measured
SEE spectra are shown in Fig.~\ref{Fig8} below, where we see
asymmetric spectra with downshifted components. We also observe a
slow temporal change of the spectrum, which indicates that the
plasma parameters in the ionosphere are slowly modified by the
large-amplitude electromagnetic wave.

In Fig.~\ref{Fig4}, we have varied the pump field $E_{\rm pump}$ and
the ionospheric length scale $L$. We define the spectral width as
the frequency difference between the high-frequency and
low-frequency points where the power has dropped 30 dB compared to
the downshifted maximum, as indicated with dashed lines in column a)
of Fig.~\ref{Fig4} for $E_{\rm pump}=1.0\times 10^{-4}$ V/m; here
the spectral width is approximately 5 kHz. We see in column a) of
Fig.~\ref{Fig4} that the spectral width is almost linearly
proportional to the amplitude of $E_{\rm pump}$. The spectra also
widens for larger ionospheric length scales, seen in column b) of
Fig.~\ref{Fig4}, and the spectral width is approximately
proportional to the square root of the length scale $L$. In
Fig.~\ref{Fig5}, we have varied the collision frequencies
$\gamma_{L}$ and $\gamma_{S}$. We here see that the width of the
spectrum increases for smaller values of $\gamma_{L}$ and
$\gamma_{s}$, and is approximately inversely proportional to the
square root of $\gamma_{L}$ and $\gamma_{s}$.

In Figs.~\ref{Fig6} and \ref{Fig7}, we have repeated the simulations
for a one order of magnitude larger pump field than in
Figs.~\ref{Fig4} and \ref{Fig5}. The qualitative results are the
same in Figs.~\ref{Fig6} and \ref{Fig7}. In Fig.~\ref{Fig6} we see
that the with of the spectrum is approximately linearly proportional
to $E_{\rm pump}$. Comparing the case $E_{\rm pump}=10^{-3}$ V/m in
column a) of Fig.~\ref{Fig6}, where the spectral with is 50 kHz with
the case $E_{\rm pump}=10^{-4}$ V/m in column a) of Fig.~\ref{Fig4},
where the spectral width is 5 kHz, we see that the linear dependence
of the spectral width on the pump amplitude holds over one decade of
pump amplitudes. Similar to the results in Figs.~\ref{Fig4} and
\ref{Fig5}, the spectra in Figs.~\ref{Fig6} and \ref{Fig7} widen for
increasing $L$ and decreasing $\gamma_L$ and $\gamma_s$, though the
downshifted parts of the spectra are slightly less sensitive to
variations in these parameters. Hence, the spectral width obeys the
approximate scaling law
\begin{equation}
  [\Delta f]_{\rm w} \simeq
  C \frac{E_{\rm pump} L^{1/2}}{\gamma_{L}^{1/2} \gamma_{S}^{1/2}},
  \label{scaling1}
\end{equation}
where $C$ is a normalization constant that depends on temperatures,
the density, etc., but not on $E_{\rm pump}$, $L$, $\gamma_{L}$ or
$\gamma_{s}$. The scaling law can also be expressed in terms of the
normalized, primed variables in Eq.~(\ref{primed}) as
\begin{equation}
  [\Delta f']_{\rm w} \simeq
  C' \frac{E_{\rm pump}' L'^{1/2}}{\gamma_{L}'^{1/2} \gamma_{S}'^{1/2}},
  \label{scaling2}
\end{equation}
where $C'$ is a numerical scaling factor that does not depend on any
physical parameters in the Zakharov system.

By the scaling of the time (\ref{scaling_t}) in the normalized
system (\ref{primed}), we have the the relation
\begin{equation}
  [\Delta f]_{\rm w}= \frac{2C_s^2\omega_{pe}}{3v_{Te}^2}[\Delta f']_{\rm w}
  \label{scaling3}
\end{equation}
between the dimensional and dimensionless spectral widths. Inserting
(\ref{scaling1}) and (\ref{scaling2}) into (\ref{scaling3}), and
eliminating $E_{\rm pump}$, $L$, $\gamma_{L}$ and $\gamma_{s}$ with
the help of (\ref{scaling_Epump})--(\ref{scaling_gs}), we find,
after some algebra, the relation
$C=[(\sqrt{6}/18)\omega_{pe}^{3/2}C_s^{5/2}e/v_{Te}^4(m_e
m_i)^{1/2}]C'$ between the dimensional and dimensionless scaling
factors. Hence, the scaling law (\ref{scaling1}) can be expressed as
\begin{equation}
  [\Delta f]_{\rm w} \simeq
  A' \frac{\omega_{pe}^{3/2}C_s^{5/2}e E_{\rm pump} L^{1/2}}{v_{Te}^4(m_im_e)^{1/2}\gamma_{L}^{1/2} \gamma_{s}^{1/2}},
  \label{scaling4}
\end{equation}
where $A'=(\sqrt{6}/18)C'$ is a numerical, dimensionless factor,
which is determined by the definition of the spectral width. As an
example, we take the third row from top of panels in
Fig.~\ref{Fig4}, where we see that the spectral width is $[\Delta
f]_{\rm w}\simeq 5 \times 10^3\,\mathrm{Hz}$, and where the
parameters are $E_{\rm pump}=10^{-4}\,\mathrm{V/m}$, $L=50\times
10^3\,\mathrm{m}$, $\gamma_{L}=\gamma_{s}=10^3\,\mathrm{s}^{-1}$,
etc. Inserting the spectral width and the parameters into
Eq.~(\ref{scaling4}), we determine the scaling factor to $A'=4$. We
have thus in Eq.~(\ref{scaling4}) a quantitative measure of the
dependence of the frequency spectrum on the physical parameters in
the ionosphere.

Returning to the experimental results in Fig.~\ref{Fig8}, we see
that the spectral width of the SEE spectrum increases slowly with
time, from approximately 40 kHz in the time intervals during the
first second, to approximately 80--90 kHz after 2--4 seconds. This
could possibly be due to an increase of the electron temperature, so
that $T_e\gg T_i$, and a resulting decrease of the ion Landau
damping $\gamma_s$ \citep{Krall} with time, which in turn would lead
to an increase of the spectral width according to our scaling law
(\ref{scaling4}).

\section{Summary}
In conclusion, we have performed a simulation study of Langmuir
turbulence with parameters relevant for the F layer in the Earth's
ionosphere. We have used a Zakharov model, including collisions and
an ionospheric plasma homogeneity, whose length scale is of the
order $10$--$100$ km. The energy source is assumed to be provided by
electromagnetic waves \citep{Thide82} in controlled ionospheric wave
injection experiments, and is modeled by a constant source term in
the Zakharov equation. We have analyzed the frequency spectra for
different sets of parameters and different altitudes relative to the
classical turning point of the Langmuir waves. By a parametric
study, we have derived a simple scaling law, which links the
spectral width of the turbulent frequency spectrum to the physical
parameters in the ionosphere. The scaling law provides a
quantitative relation between the physical parameters (temperatures,
electron number density, ionospheric length scale, etc.) and the
observed frequency spectrum. An application of our results is to
compare our scaling law with SEE spectra, with the assumption that
the processes creating the SEE reflect the spectral structure of the
turbulent $E$ field via mode conversion of electrostatic waves to
electromagnetic radiation.

{\bf Acknowledgment} This work was supported financially by the
Swedish Research Council (VR).



\begin{thebibliography}{}

\bibitem[{\it Carozzi et al.}(2002)]{Carozzi02} Carozzi, T. D., B. Thid\'e, S. M. Grach, T. B.
Leyser, M. Holz, G. P. Komrakov, V. L. Frolov, and E. N. Sergeev
(2002) Stimulated electromagnetic emissions during pump frequency
sweep through fourth electron cyclotron harmonic, \textit{J.
Geophys. Res.}, {\it 107}(A9), 1253,
doi:10.1029/2001JA005082.\vspace{0.2cm}

\bibitem[{\it Cros et al.}(1991)]{Cros01} Cros, B., J. Godiot, G. Matthieussent, and A. H\'eron (1991)
Laboratory simulation of ionospheric heating experiment,
\textit{Geophys. Res. Lett.}, {\it 18}(8), 1623--1626.\vspace{0.2cm}

\bibitem[{\it DuBois et al.}(1991)]{DuBois91} Dubois, D. F., H. A. Rose, and D. Russell (1991) Coexistence
of parametric decay cascades and caviton collapse at subcritical
densities, \textit{Phys. Rev. Lett.}, {\it 66},
1970--1973.\vspace{0.2cm}

\bibitem[{\it DuBois et al.}(1993)]{DuBois93}
Dubois, D. F., A. Hanssen, H. A. Rose, and D. Russell  (1993)
 Space and time distribution of HF excited Langmuir turbulence in the ionosphere: Comparison of theory and experiment
\textit{J. Geophys. Res.}, {\it 98}(A10) 17543--17568.\vspace{0.2cm}

\bibitem[{\it Forme}(1993)]{Forme93}  Forme, F. R. E. (1993) A new interpretation on the origin of
enhanced ion-acoustic fluctuations in the upper atmosphere,
\textit{Geophys. Res. Lett.}, {\it 20}, 2347–-2350.\vspace{0.2cm}

\bibitem[{\it Foster et al.}(1988)]{Foster88} Foster, J. C., C. del Pozo, K. Groves, and
J.-P. St. Maurice (1988) Radar observations of the onset of current
driven instabilities in the topside ionosphere, \textit{Geophys.
Res. Lett.}, {\it 15}, 160–-163.\vspace{0.2cm}


\bibitem[{\it Frolov et al.}(2004)]{Frolov04} Frolov, V. L., E. N. Sergeev, G. P. Komrakov,
P. Stubbe, B. Thid\'e, M. Waldenvik, E. Veszelei, T. B. Leyser
(2004) Ponderomotive narrow continuum (NC$_p$) component in
stimulated electromagnetic emission spectra, \textit{J. Geophys.
Res.}, {\it 109}, A07304/1--21.\vspace{0.2cm}

\bibitem[{\it Guio and Forme}(2006)]{Guio06} Guio, P., F. Forme (2006) Zakharov simulations of Langmuir
turbulence: Effects on the ion-acoustic waves in incoherent
scatteroing, \textit{Phys. Plasmas}, {\it 13}, 122902.\vspace{0.2cm}

\bibitem[{\it Hansen et al.}(1992)]{Hansen92}
 Hanssen, A., E. Mj{\o}lhus, D. F. Dubois and H. A. Rose  (1992)
Numerical test of the weak turbulence approximation to ionospheric
Langmuir turbulence, \textit{J. Geophys. Res.}, {\it 97}(A8) 1992,
12073--12091.\vspace{0.2cm}

\bibitem[{\it Kim et al.}(1974)]{Kim74} Kim, H. C., R. L. Stenzel, and A. Y. Wong (1974)
Development of "cavitons" and trapping of rf field, \textit{Phys.
Rev. Lett.}, {\it 33}(15), 886--889.\vspace{0.2cm}

\bibitem[{\it Krall and Trivelpiece}(1973)]{Krall} Krall, N. A., and A. W. Trivelpiece (1973) {\em Principles of
Plasma Physics} (McGraw-Hill, New York).\vspace{0.2cm}

\bibitem[{\it Leyser}(2001)]{Leyser01} Leyser, T. B.  (2001) Stimulated electromagnetic emissions by
high-frequency electromagnetic pumping of the ionospheric plasma,
\textit{Space Sci. Rev.}, {\it 98}, 223–-328.\\[0.1cm]

\bibitem[{\it Mj{\o}lhus}(1990)]{Mjolhus90} Mj{\o}lhus, E. (1990), On linear conversion in a
magnetized plasma, \textit{Radio Sci.} \textit{25}(6),
1321--1339.\vspace{0.2cm}

\bibitem[{\it Mj{\o}lhus}(2003)]{Mjolhus03} Mj{\o}lhus, E., E. Helmersen, and D. F. DuBois (2003)
Geometric aspects of HF driven Langmuir turbulence in the
ionosphere, \textit{Nonlin. Proc. Geophys.}, {\it 10},
151-–177.\vspace{0.2cm}

\bibitem[{\it Morales and Lee}(1974)]{Morales74} Morales, G. J. and Y. C. Lee (1974)
Ponderomotive-force effects in a nonuniform plasma, \textit{Phys.
Rev. Lett.}, {\it 33}(17), 1016--1019.\vspace{0.2cm}

\bibitem[{\it Morales and Lee}(1977)]{Morales77} Morales, G. and Y. C. Lee (1977)
Generation of density cavities and localized electric fields in
nonuniform plasma, \textit{Phys. Fluids.}, {\it 20},
1135-1147.\vspace{0.2cm}

\bibitem[{\it Nicholson et al.}(1984)]{Nicholson84} Nicholson, D. R., G. L. Payne, R. M. Downie,
and J. P. Sheerin  (1984) Solitons versus parametric instabilities
during ionospheric heating, \textit{Phys. Rev. Lett.}, {\it 52},
2152-2155.\vspace{0.2cm}

\bibitem[{\it Robinson}(1997)]{Robinson97} Robinson, P. A. (1997) Nonlinear wave collapse and strong
turbulence, \textit{Rev. Mod. Phys.}, \textit{69},
507--574.\vspace{0.2cm}

\bibitem[{\it Sedgemore-Schulthess and St. Maurice}(2001)]{Sedgemore01} Sedgemore-Schulthess, F. and J.-P. St.
Maurice (2001) Naturally enhanced ion-acoustic spectra and their
interpretation, \textit{Surv. Geophys.}, {\it 22}, 55–-92,
doi:10.1023/A:1010691026863.\vspace{0.2cm}

\bibitem[{\it Stubbe et al.}(1984)]{Stubbe84} Stubbe, P., H. Kopka, B. Thid{\'e}, and H.
Derblom (1984) Stimulated electromagnetic emission: A new technique
to study the parametric decay instability in the ionosphere,
\textit{J. Geophys. Res.}, {\it 89}, 7523--7536.\vspace{0.2cm}

\bibitem[{\it Thid\'e et al.}(1982)]{Thide82} Thid\'e, B., H. Kopka and P. Stubbe (1982)
\textit{Phys. Rev. Lett.}, {\it 49}, 1561--1564.\vspace{0.2cm}

\bibitem[{\it Thid\'e et al.}(1989)]{Thide89} Thid{\'e}, B., {\AA}. Hedberg, J. A. Fejer, M.
Sulzer (1989) First observation of stimulated electromagnetic
     emission at Arecibo, \textit{Geophys. Res. Lett.}, {\it 16}, 369--372.\vspace{0.2cm}

\bibitem[{\it Thid\'e}(1990)]{Thide90} Thid{\'e}, B. (1990) Stimulated scattering of large
amplitude waves in the ionosphere: Experimental results,
\textit{Phys. Scr.}, {\it T30}, 170--180.\vspace{0.2cm}

\bibitem[{\it Wong et al.}(1987)]{Wong87} Wong, A. Y., T. Tanikawa, and A. Kuthi (1987)
Observations of ionospheric cavitons, \textit{Phys. Rev. Lett.},
{\it 58}(13), 1375--1378.








\end{thebibliography}
\end{document}